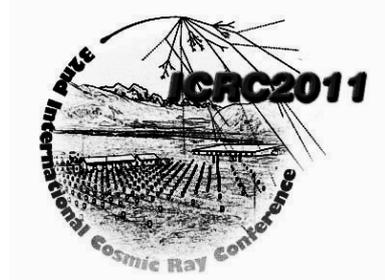

# Status of the VERITAS Upgrade

D. B. KIEDA[1] FOR THE VERITAS COLLABORATION[2]
[1]Department of Physics and Astronomy, University of Utah, Salt Lake City, UT 84112 USA
[2]see J. Holder et al. (these proceedings)
contact: dave.kieda@utah.edu

**Abstract:** The VERITAS gamma ray observatory (Amado, AZ, veritas.sao.arizona.edu) uses the Imaging Atmospheric Cherenkov Technique (IACT) to study sources of Very High Energy (VHE: E > 100 GeV) gamma rays. Key science results from the first three years of observation include the discovery of the first VHE emitting starburst galaxy, detection of new Active Galactic Nuclei (AGN), SuperNova Remnants (SNR), gamma ray binaries as well as strong limits on the emission of VHE gamma rays from dark matter annihilation in dwarf galaxies. In April 2010, VERITAS received funding to upgrade the photomultiplier tube cameras, pattern triggers, and networking systems in order to improve detector sensitivity, especially near detection threshold (E ~ 100 GeV). In this paper we describe the status of the VERITAS upgrade and the expected improvements in sensitivity when it is completed in summer 2012.

**Keywords:** gamma ray astronomy, instrumentation, VERITAS, Imaging Atmospheric Cherenkov Technique

## 1 Introduction

The VERITAS Observatory [1, 2], located near Amado, AZ, employs an array of four 12-m diameter telescopes to study GeV/TeV $\gamma$-rays using the Imaging Atmospheric Cherenkov technique. Each telescope employs a 499-pixel photomultiplier (PMT) imaging camera at the focal plane, and 500M Samples per Second (MSPS) Flash ADC (FADC) readout electronics to record the Cherenkov Images observed by each telescope. Since its commissioning in 2007, VERITAS has greatly expanded the catalog of Northern Hemisphere VHE $\gamma$-rays sources, including the discovery of new extragalactic blazars [3, 4, 5], starburst galaxies [6], galactic objects including supernova remnants [7, 8, 9], and pulsar emission from the Crab extending above 100 GeV [10, 11]. VERITAS has also set new constraints the presence of Minimal SuperSymmetric Model (MSSM) dark matter annihilation in dwarf galaxies [12].

In parallel with the science observation program, the VERITAS Collaboration has worked to continually improve the sensitivity of the observatory. A new mirror alignment technique has been developed and employed [13]; this technique substantially improves the point spread function and speed of mirror alignment. In summer 2009, VERITAS telescope #1 was disassembled, refurbished, and redeployed to a new location east of the other three telescopes (Figure 1), creating a more favorable array geometry. Improvements in analysis techniques and calibration have also led to gains in detector sensitivity. The combination of these improvements has reduced the time to observe a 1% Crab-flux point source (E > 100 GeV) from approximately 46 hours (in 2007) to approximately 25 hours in 2010 (Figure 2).

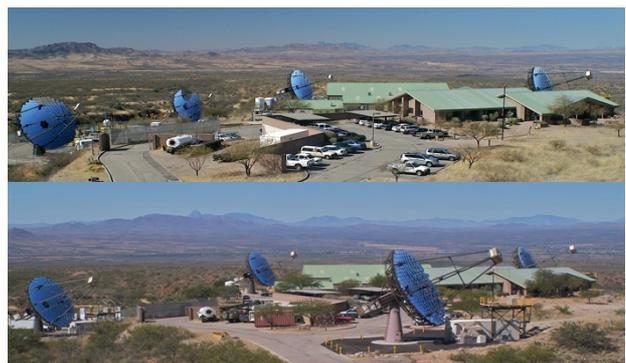

Figure 1: VERITAS array telescope arrangement. Upper Photo: 2007-July 2009; Lower Photo: Aug 2009-present

Additional major gains in VERITAS sensitivity can be achieved through improvements in analysis as well as hardware improvements that will improve PMT quantum efficiency (QE) and reduce background event rates. In 2010, the VERITAS collaboration began a 3-year upgrade project to replace the existing Photonis XP2970 PMTs (18-22% QE) with Hamamatsu R10560 PMTs (32-34% QE) [14]. This VERITAS upgrade will also replace the existing Level 2 Pattern (telescope-level) trigger with a higher speed, FPGA-based pattern trigger system with a narrower coincidence window [15]. Finally, the VERITAS upgrade will improve inter-telescope networking and communications, and will add the capability of using the VERITAS Observatory for high-speed Optical Monitoring (OM) and Stellar Intensity Interferometry (SII) [16, 17]. In this paper, we will describe the details of the PMT mechanical design and the network upgrade. The PMT tests and qualifications [14] and the FPGA-based pattern trigger [15] are described in separate papers at this conference.

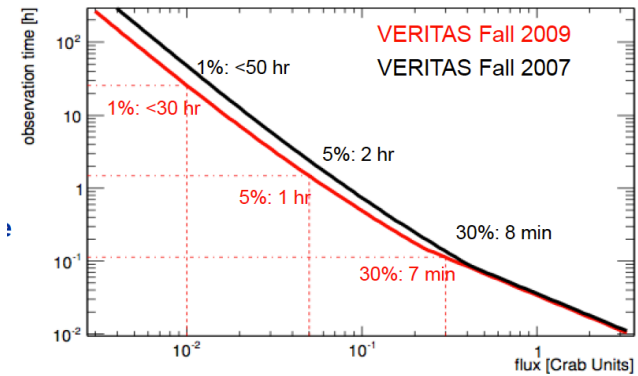

Figure 2: Improvements in VERITAS sensitivity (2007-2009)

## 2   VERITAS high-QE (hQE) Pixel Design

The components of the VERITAS hQE pixel design are shown in Figure 3. Each pixel contains a high QE PMT and preamplifier combined with mechanical mounting tubes to hold the PMT and to mount the pixel to the focal plane.

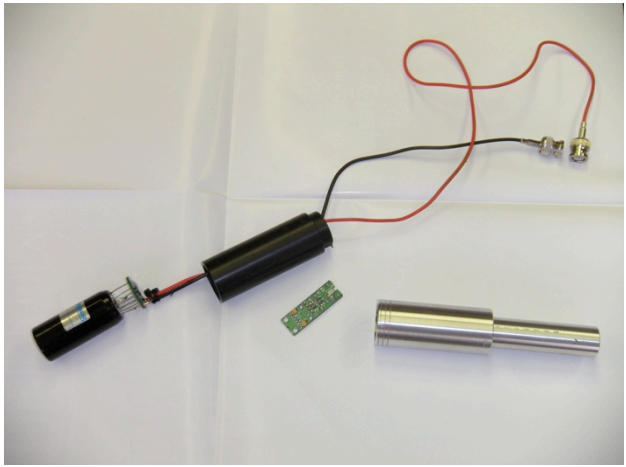

Figure 3: VERITAS hQE Pixel Components. From left to right: Hamamatsu R10560-100-20 High QE PMT, Delrin mounting tube, preamplifier, aluminum mounting tube.

### 2.1 Photomultiplier Tubes and Testing

The Hamamatsu R10560-100-20 super-bialkali PMT was chosen after careful evaluation of PMT gain-voltage, timing, cathode uniformity, spectral response and afterpulsing [14]. This PMT has a slightly smaller diameter than the currently used Photonis XP2970 (26.2 mm vs. 29 mm), but since the VERITAS cameras already use Winston cones to improve photon collection at the focal plane, the new PMTs will not suffer any geometrical loss of photon collection area.

The photomultiplier tubes are being delivered from the manufacturer at a rate of 250 PMTs/month, commencing May 2011. The Purdue University group performs acceptance testing of the individual PMTs by placing each PMT under high voltage (HV) for approximately 1 hour, and then characterizes and calibrates each PMT by acquiring a single photon spectrum for 8 different HV settings. This allows derivation of the Gain-Voltage relationship as well as measurement of the absolute gain using the single photoelectron peak. A typical single photon spectrum is displayed in Figure 4 for one of the first prototype R10560-100-20 PMTs. All data for each PMT is stored in an online database. The testing is done with batches of 16 PMTs in parallel; full testing of a 16 PMT batch takes approximately 3 hours. In addition, a random sample of 10% of the PMTs are tested for cathode uniformity and quantum efficiency vs. wavelength. Independent tests for after-pulsing fraction, absolute gain and photoelectron detection efficiency are also performed for a random 10% subset of the PMTs by VERITAS collaborators at UC Santa Cruz.

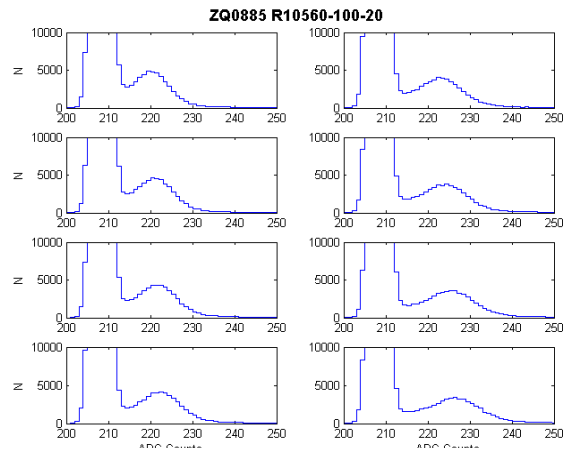

Figure 4: The single photon spectrum for voltages from 1385 V (upper left) to 1490 V (lower right) for the R10560-100-20 PMT with serial # ZQ0885.

### 2.2 Mechanical Design

The new VERITAS hQE pixels have been designed to be `plug-compatible' with the existing VERITAS pixels. This design was chosen so as to allow rapid pixel substitution with minimal risk of incompatibility with other camera and electronics components. The mechanical design was also required to meet environmental tests for mechanical stability and usability over a wide temperature range (0° F to 140° F). The new two-piece mechanical design (Figure 5) was improved over the existing design to allow easier/more accurate pixel fabrication without the use of an alignment jig and allow for possible field dis-assembly/re-assembly.

The total weight of the hQE pixels is reduced in the new design. Although the mounting tube is slightly heavier than the previous design, the PMT and bleeder chain are substantially lighter, so the overall weight of a single hQE pixel is 7.7% less than a current VERITAS pixel.

## 2.3 Preamplifier Design

The design of the hQE pixel maintains the identical power, signal, and charge injection polarities and connectors with the existing pixels, and the design was required to have similar/lower power consumption to the existing pixel design. In order to meet these requirements with minimal design risk, the hQE pixels use the same preamplifier design that is employed in the current VERITAS pixels [15] (Figure 3). The hQE pixel cable shielding uses a nylon fabric `spaghetti' to improve flexibility and reduce weight compared to the previous design.

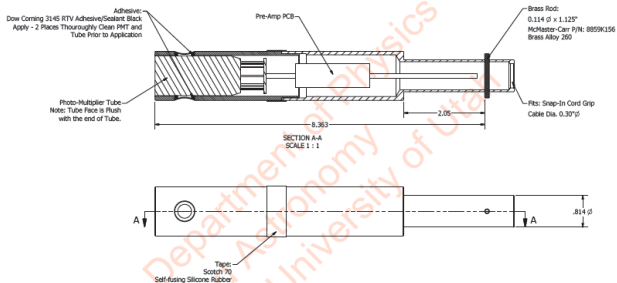

Figure 5: Two-piece hQE Pixel mechanical design

## 2.4 Pixel Stability Tests

During October 2010, 14 prototype hQE pixels were installed in VERITAS Telescope #3 in order to assess the long-term stability of the hQE pixel performance. Figure 6 illustrates the stability of the hQE pixels compared to the average T3 (non-hQE) pixel gain, tested over a six-month field deployment. The data indicates the prototype hQE pixel gain stability and performance is suitable for use in the full VERITAS upgrade.

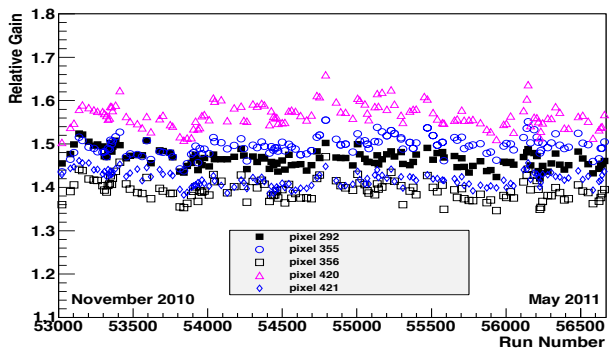

Figure 6: Observed 6-month gain stability of five prototype hQE pixels. The measured gain (vertical axis) is relative to the average (non-hQE) pixel gain.

## 3 Optical Monitor (OM)/SII System

The VERITAS upgrade includes the special instrumentation of a single, dedicated high-speed pixel at the center of each camera for OM/SII measurements. The pixel will employ a removable narrowband filter ($\Delta\lambda/\lambda \approx 10^{-4}$) that is used for intensity interferometry applications. The pixel is instrumented with a high-speed National Instruments PXIe-7965R FPGA module (FlexRIO, Virtex 5 SX95T) combined with a NI-5761 digitizer (4-channel, 250 MSPS/s, 14 bit resolution) that can continuously stream the photomultiplier tube signal at 500 MSPS/8-bit resolution to a 12 TByte NI-HDD 8265 disk array. The PMT data stream can be streamed uninterrupted for up to six hours. The data is analyzed during the next day for periodic signals such as optical pulsar signals. Cross-correlation of optical intensity data between telescopes can be used to perform intensity interferometry analysis that can extract a measurement of the stellar radii of nearby bright, hot stars.

## 4 Network Upgrade

During the first two years of operation, the VERITAS observatory used a single pair of multi-mode (62.5/125 μ) fiber optic links to send individual telescope data to a central data `harvester' for event building and archiving, as well as control and status information to the central control building. These fiber-optic data links were limited to 1 GB/s bandwidth, and possessed no redundancy, so that a failed fiber optic cable or transceiver would result in complete loss of data from an individual telescope.

In Summer 2010, all fiber-optic communication links were upgraded to six-pair of single-mode (50/125 μ) optical fiber; each fiber is capable of transmitting up to 10 GB/sec (Figure 7). The network switches and routers were upgraded to handle LACP protocol, which allows splitting a single data link over multiple physical fiber-optic links, thereby improving bandwidth as well as providing redundancy that provides real-time protection against failure of an single fiber optic cable or transceiver. Because of this, the network bandwidth for individual telescopes to the central harvester has now. We anticipate upgrading the network switches and transceivers in summer 2012 to bring the full single telescope bandwidth up to 20 MB/s. This will potentially allow full data streaming of the SII system as well as possible implementation of higher level triggering systems that may reduce cosmic ray background and potentially increase sensitivity to low energy (E < 100 GeV) gamma ray events.

## 5 Expected Sensitivity Gain

The VERITAS upgrade will have its most significant impact on the ability to detect and reconstruct gamma rays at energies below 200 GeV. Figure 8 illustrates the expected gain in detection are as a function of primary gamma ray energy for various improvements in PMT quantum efficiency. The hQE pixels (Figure 3) have demonstrated a 35-50% increase in photon sensitivity relative to the existing VERITAS PMTS, and so we expect the gain in effective are to be near the upper limits of the plotted points.

## 6 Schedule

The VERITAS upgrade project began in Summer 2010, and individual components of the upgrade will be deployed during a three-year construction timeline. During Summer 2010, the first phase of the network upgrade was completed. During this time, the FPGA pattern trigger was built, and testing of the triggering system took place in Fall 2010. The FPGA pattern trigger software and control system is currently in development, and will be field-tested in June 2011. The full FPGA-pattern trigger upgrade will be installed and tested after successful completion of this testing.

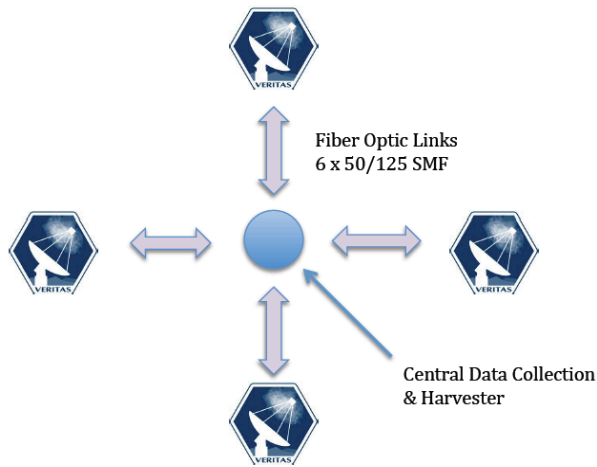

Figure 7: VERITAS communication network configuration.

The OM/SII system is currently under development. As of May 2011, a 500 MSPS/channel, two-channel, 8-bit resolution, continuous streaming data acquisition system has been built and tested in the lab. Initial tests of the system have employed two Hamamatsu R10560-100-20 high QE PMTs arranged in a fiber-optic intensity interferometer with a narrow line blue emission line discharge tube. The system has demonstrated cross-channel correlation to autocorrelation ratio less than $10^{-4}$, which is sufficient for measurement of stellar radii of nearby bright stars using stellar intensity interferometry. Field tests of a prototype SII system, using a pair of 3m diameter telescopes separated by 23 m, is scheduled for Summer 2011.

After successful completion of these tests, the design of the full 4-telescope OM/SII system will be completed. The OM/SII system will be installed in at the VERITAS telescope site during Summer 2012, during the same time the hQE Pixels are being installed and tested. We anticipated upgrading the networking switches during Summer 2012 in order to accommodate 20 Gbit/sec connectivity between each telescope and the central trailer. This higher bandwidth will allow fast streaming of the full OM/SII data stream from each telescope to the central data recording trailer.

Initial PMT testing and selection was completed by November 2010, and an order for 2200 PMTs was placed shortly afterwards. We received the first shipment of 191 high-QE PMTs from Hamamatsu in May 2011, and the remaining PMTs will be delivered at a rate of 250/month.

The PMT testing and hQE pixel assembly will follow at the same rate of 250 hQE pixels/month. We anticipate the last PMTs will be delivered in January 2012, and the hQE pixel fabrication and tests will be completed by April 2012. Installation of the hQE pixels for all four telescopes will take place during Summer 2012, with first light on the upgraded VERITAS Observatory scheduled for the beginning of the 2012-2013 observing season.

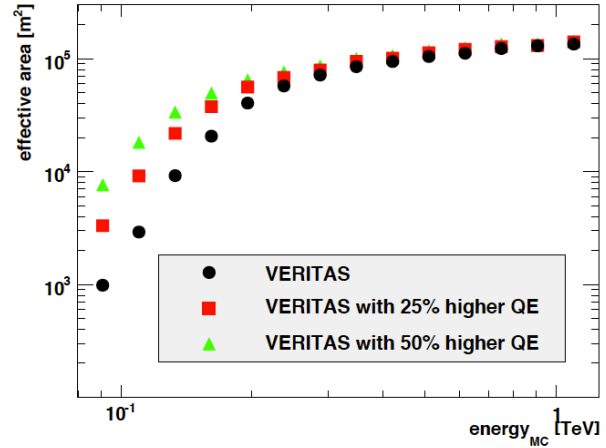

Figure 8: Simulated increase in VERITAS effective detection area for various improvements in PMT quantum efficiency.

## Acknowledgements

This research is supported by grants from the US Department of Energy, the US National Science Foundation, and the Smithsonian Institution, by NSERC in Canada, by Science Foundation Ireland, and by STFC in the UK. We acknowledge the excellent work of the technical support staff at the FLWO and the collaborating institutions in the construction and operation of the instrument.